\documentclass{aastex}          
\usepackage{spr-astr-addons}    

\usepackage[cp866]{inputenc}
\usepackage[english]{babel}
\usepackage{graphicx}
\usepackage{amsmath}
\usepackage{amssymb}
\usepackage{epsf}
\usepackage{bm}

\def\la{\;\raise0.3ex\hbox{$<$\kern-0.75em\raise-1.1ex\hbox{$\sim$}}\;}
\def\ga{\;\raise0.3ex\hbox{$>$\kern-0.75em\raise-1.1ex\hbox{$\sim$}}\;}

\shorttitle{Periodicity in  distribution}
\shortauthors{Ryabinkov,  Kaminker}

\begin{document}

%
\title{Large-scale periodicity in the distribution   \\
of QSO absorption-line systems}

%
\author{A.~I.~Ryabinkov  and  A.~D.~Kaminker}  	 
\affil{Ioffe Physical-Technical Institute,
Politekhnicheskaya 26, 194021 St.~Petersburg, Russia}

\email{calisto@rbcmail.ru,  kam@astro.ioffe.ru}
%


\begin{abstract}
The spatial-temporal
distribution of absorp\-tion-line systems (ALSs) 
observed in QSO spectra within the cosmological 
redshift interval $z=$0.0--4.3 is  investigated
on the base of our updated
catalog of absorption systems. 
We consider so called metallic systems including 
basically lines of heavy elements.
The sample of the data displays regular 
variations  
(with amplitudes $\sim 15$ -- $20\%$) 
in the $z$-distribution of ALSs 
as well as in the 
$\eta$-distribution,
where $\eta$ is a 
dimensionless line-of-sight comoving distance,
relatively to  smoother dependences.   
The  $\eta$-distribution reveals the periodicity 
with period $\Delta \eta = 0.036 \pm 0.002$, 
which corresponds to 
a spatial characteristic scale $(108 \pm 6)~h^{-1}$~Mpc   
or (alternatively)   
a temporal interval $(350 \pm 20)~h^{-1}$~Myr 
for the  $\Lambda$CDM cosmological model. 
We discuss a possibility of a spatial 
interpretation of the results treating 
the pattern obtained as a trace of 
an order imprinted on the galaxy clustering 
in the early Universe. 
\end{abstract}
   

%
\section{Introduction}
\label{s:intro}
In our previous papers
(e.g.,  \citeauthor*{krv00} \citeyear{krv00},  
\citeauthor*{rkv07} \citeyear{rkv07},\ 
hereafter Paper~I) we studied
the spatial-temporal distribution
of absorption-line systems (ALSs)
imprinted in spectra of quasars (QSOs).
We regard that the redshifts $z_a$ 
of absorption lines
are cosmological, and detected ALSs
are associated with ionized gas in intervening
galaxies or clusters of galaxies at cosmological
distances along lines of sight to original QSOs. 
Thus  a representative 
sample of ALSs is considered to  trace 
the distribution of matter 
between the observer and  QSOs,
as well as physical processes occurred 
in different epochs
of the cosmological evolution.
   
In Paper~I  we explored 2003 ALSs  
in the redshift range $z=0.0$--3.7
from our catalog of absorption systems
\citep{rkv03}  and restricted ourselves
by consideration of so called metallic   
systems including lines of heavy elements. 
ALSs registered in a spectrum of 
a quasar might comprise  
up to 20 -- 30 absorption lines 
predominantly  within an interval  
$\sim$ 3000 -- 8000 \AA.  It was shown 
that the $z$-distribution of ALSs displays
a pattern of alternating maxima (peaks)
and minima (dips) which
are statistically significant
against a smoother dependence (trend).
The positions of the peaks and dips 
obtained for different hemispheres 
on the sky turned out to be
independent (within statistical uncertainties)
of observation directions.
Moreover, it was shown that 
the sequence of peaks and dips 
reveals a certain regularity.
The power spectrum   
calculated with using a 
rescaling function $\tau(z)$
(Eqs.~(11) and (12) of Paper~I)   
displays the peak  
at the significance level exceeding $4 \sigma$ 
relatively to the hypothesis 
of the uniform distribution of ALSs 
along the axis $\tau(z)$.
    
In this paper we present the results 
of slightly  extended statistical analysis
performed  by similar methods
as we used in Paper~I but in a new bearing. 
The principle point of this paper is the  
analysis  of the ALSs distribution
in the  comoving coordinate system 
(CS). Such a representation enables us 
to find the periodicity of the distribution
and treat it as 
a trace of partly ordered spatial 
structures. 
\begin{figure*}[t]   
\epsfxsize=.75\linewidth
\epsffile[-50 200 530 770]{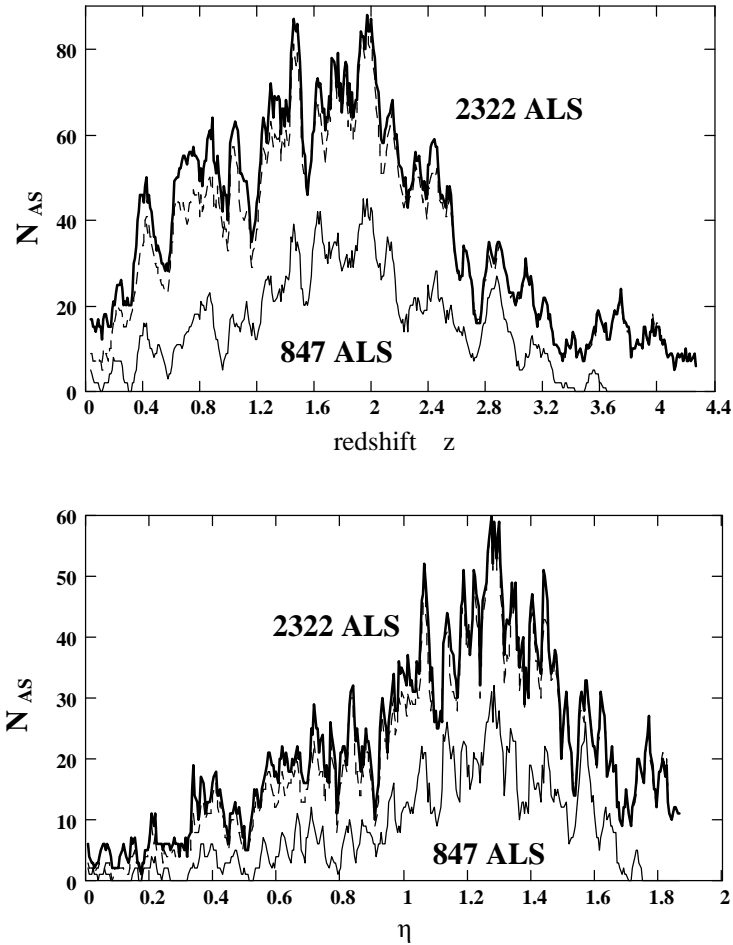}
\caption{
\footnotesize
{\it Upper panel:} 
Sliding-average $z$-distributions
of absorption-line systems (ALSs) observed in QSO spectra 
within the  redshift interval $z=0.0$ -- 4.30
with an averaging bin $\Delta_{\rm z} = 0.071$ and 
a step of its consecutive shift $\delta_{\rm z} = 0.01$; 
a velocity interval of ALS 
associations (along each line of sight)
into a single system $z_i$ (see text)
is chosen as 
$\delta v = 500$km~s$^{-1}$;
thin solid lines -- 847 systems are taken      
from the catalog by   
\cite{jhb91},    
thin dashed lines -- 2003 systems --      
from the catalog by \cite{rkv03},
thick lines -- 2322 systems -- from the
same but updated catalog.
{\it Lower panel:} 
same as in the upper panel 
but for corresponding  
$\eta$-distributions (see text)
within the appropriate interval 
$\eta=0.0$ -- 1.87
with an averaging bin $\Delta_{\eta} = 0.018$ and 
a step of shifting $\delta_{\eta} = 0.005$.
}
\label{AS91-09}
\end{figure*}
 
The upper panel of Fig.\ \ref{AS91-09} 
demonstrates three $z$-distributions  
$N_{\rm AS}(z)$ of ALSs
containing absorption lines of heavy elements
within the redshift
interval $z=0.0$ -- 4.3. 
One of them is obtained
using 847 systems from 
the catalog by  \cite{jhb91},
the second one is based on 
2003 systems 
registered in the spectra of 661 QSOs 
(emission redshifts 
$z_e=$0.29 -- 4.9) from the
catalog by \cite{rkv03}.  
The third one is the $z$-distribution
of 2322 ALSs registered in the spectra 
of 730 QSOs ($z_e=$0.035 -- 5.01)
from the same but  
updated catalog. All three samples 
of ALSs represent the same types
of metallic systems including  
single lines or resonance
doublets of ions: C~II--IV, Mg~II, Si~II--IV,
N~V, Al~II--III, Fe~II etc. 
They exclude ALSs consisting 
only of neutral hydrogen lines as well as 
containing damped Ly$\alpha$ 
absorption systems (DLA).
      
Following to \cite{krv00} 
we use the procedure of smoothing away
an initial sample of ALS redshifts
to reveal possible large scale variations 
and suppress small scale clustering 
of the ALSs.   
All redshifts $z_j$ 
registered in a spectrum of a certain QSO
and fallen into the velocity interval
$\, \delta v =$500 km~s$^{-1}$
are treated as a single absorption system
with an averaged redshift
$z_a=(\sum_{j=1}^{n_a} z_j)/n_a$,
where $n_a$ is a number of redshifts 
included in the group.  
Three distributions in the upper panel 
of Fig.~\ref{AS91-09}
are obtained using so-called
sliding-average approach  which represents a set of
consecutive displacements of the averaging bin
$\Delta_{\rm z}=0.071$  along $z$-axis
with a step $\delta_{\rm z}=0.01$.

The lower panel of Fig.\ \ref{AS91-09} 
represents the same three distributions
but calculated relatively to a dimensionless 
line-of-site comoving distance $\eta(z)$
(e.g., \citeauthor{h93}  \citeyear{h93}, 
\citeauthor{khs97}  \citeyear{khs97}, 
\citeauthor{h99}  \citeyear{h99}):
\begin{equation}
  \eta(z_i)  =   \int_0^{z_i} 
      {1 \over \sqrt{\Omega_{\rm m} (1+z)^3 +
       \Omega_{\rm \Lambda}}} \, \,  {\rm d}z ,
%
\label{eta}
\end{equation}
where $i$ is a numeration of all ALSs
$i=1,2, ... {\rm N}_{\rm tot}$ 
sampled  as described above, 
N$_{\rm tot} = 847,\,  2003$, and  ~2322, respectively. 
We use the $\Lambda$CDM-cosmological model with  
the dimensionless density parameters
$\Omega_{\rm m}=0.23$ and
$\Omega_{\rm \Lambda}=1-\Omega_{\rm m}=0.77$.   
  
A comparison of the $z$-distributions
reveals similar patterns of the peaks and dips 
with rather small ($\sim 15 - 20\%$) amplitude 
of variations relatively
to smoother dependences.    
The positions of the majority of  peaks and dips
remain the same after the extension of statistics
and some of them become more significant (see Paper~I).
We have similar situation 
with  $\eta$-distributions, although
there is noticeable
lack of statistics  at $\eta \lesssim 0.6$.  

The main accent of the present paper is
placed on a search for a periodicity 
of the ALSs radial distribution 
with respect to the  variable $\eta$.  
In Section\ 2 we focus on the periodicity
of the $\eta$-distribution. 
In order to avoid
sensitivity of our results
to the choice of an averaging bin and
to the procedure of a trend elimination, 
which are mutually dependent, 
we employ  in Sections\ \ref{s:period} and \ref{s:MgII}   
a point-like statistical technique.    
In Section\ 3 
we examine the periodicity of the one-dimensional
correlation function calculated in
the comoving CS. In Section\ 4 
we estimate a trace of the same periodicity 
in the distribution of resonance
absorption doublets Mg~II 
based on the data of 
the Sloan Digital Sky Survey (SDSS).   
Conclusions and discussion 
of the results in application
to the Large-Scale Structure
are represented in Section\ 5.
Two toy models for partly ordered 
simple-cubic lattice 
displaying similar periodicities are 
discussed in Appendix.

\section{Periodicity of  $\eta$-distribution}
\label{s:period}

In this Section we employ so called 
point-like statistical approach at which 
a sequence of points 
$\eta_i \equiv \eta(z_i)$
calculated with use of  Eq.~(\ref{eta}) 
is analysed.  
To sample ALSs 
for this analysis we use   
hereafter the same averaging 
velocity interval 
$\delta v=500$km~s$^{-1}$
for the association 
of ALSs
into a single redshift 
$z_i$  as in Fig.\ \ref{AS91-09}. 
Additionally,  we exclude  
all ALSs with redshifts $z_a$   
belonging to 
a region associated with  
their host  QSO,  i.e.,  at 
$|z_e - z_a|/(1+z_e) \leq  \Delta v /c$,   
where  
$z_e$ is the emission redshift
of QSOs  and $\Delta v$ is
the minimal velocity shift 
(along a line of sight)
adopted for the sample of ALSs.  
In this paper we choose
$\Delta v = 1100$ km~s$^{-1}$,
thus the actual sample for our analysis
consists of  N$_{\rm tot}=2167$ systems. 

To verify the periodicity of the $\eta$-distribution
we calculate a power spectrum 
for the whole sequence 
of points $\eta_i$: 
%
\begin{eqnarray}
         {\rm P}({\rm k}) & =  &	   	  	
	 {1 \over {\rm N}_{\rm tot}}
                          \left\{ \left[
          \sum_{i=i_0}^{{\rm N}_{\rm tot}} \cos \left(
         {2\pi {\rm k}\,  \eta_i  \over \hat{\eta}}
              \right) \right]^2  \right.
\nonumber       \\
                 & + &	       	         
      \left.  \left[ \sum_{i=i_0}^{{\rm N}_{\rm tot}}
             \sin \left(
       {2\pi {\rm k}\,  \eta_i  \over \hat{\eta}}
              \right) \right]^2 \right\},
\label{P_k}    
\end{eqnarray}
%
where $i_0 \geq 1$ corresponds to a lowest point 
of the  sequence $\eta(z_i)$, 
$\hat{\eta}$ 
$=\eta (z_{{\rm N}_{\rm tot}}) - \eta_{i_0}$
is the whole interval
under consideration, 
k is a harmonic number. 
A periodicity  yields a peak in the
power spectrum ${\cal P}= {\rm max}$ (P(k)) with a
confidence probability
\begin{equation}
   \beta = [1- \exp(-{\cal P})],
\label{beta}
\end{equation}
which is defined with respect
to the hypothesis of the Poisson distribution
of  $\eta_i$. 

\begin{figure*}[t]   
\epsfxsize=.7\linewidth
\epsffile[-100 160 530 830]{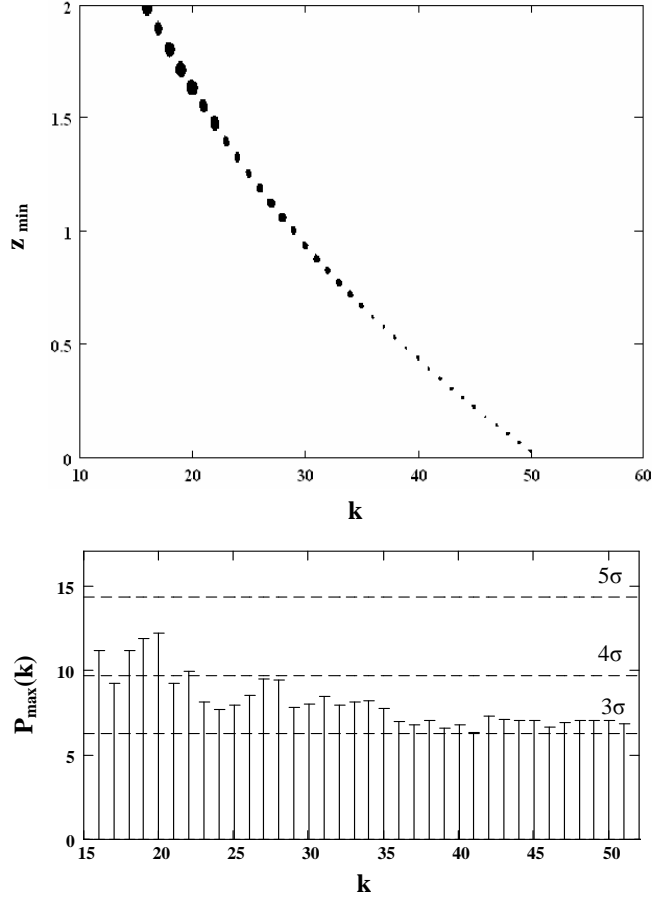}
\caption{
\footnotesize
The set of peaks of the  power spectra P$_{\rm max}$(k)
calculated according to Eq.~(\protect{\ref{P_k}})  
in a wide region of harmonic numbers k
at the significance  $\geq 3 \sigma$;
intervals of $z$  (or $\eta$)
are limited from below by $z_{\rm min}$ 
($\eta_{\rm min}$)   
shifted successively upward  along an 
interval $0.0 \leq z_{\rm min} \leq 2.0$ 
with a step  $\delta z_{\rm min}=0.01$. 
The whole sample contains 2167 values $\eta_i$  
(see text). 
All visible peaks correspond to the same period 
$\Delta \eta = 0.036 \pm 0.002$.
{\it Upper panel} indicates positions 
of the main peaks on the     
$(z_{\rm min}, {\rm k})$-plane;  
dimensions of filled regions qualitatively 
denote the values (amplitudes) of the 
peaks, smallest regions (points) 
symbolize $3 \sigma$ significance level. 
{\it Lower panel} plots the values  
of the highest-peaks P$_{\rm max}$(k) 
versus the integer number 
k corresponding to 
the regions marked on the upper panel.
The horizontal dash lines
specify the significance levels  
$3\sigma$, $4\sigma$, and $5\sigma$
calculated with use of
Eq.~(\protect{\ref{beta}})  
}
\label{Pmax}
\end{figure*}
 
Following  Eq.~(\ref{P_k}) 
we performed additionally 
numerous calculations  
of the power spectra  ${\rm P}$(k) 
for different 
realizations of randomly distributed points 
$x_i$ ($1 \leq i \leq {\rm N}_{\rm tot}$) 
within  intervals ${\rm D}x$.      
We assumed that     
density of points
$x_i$ obey to
Gaussian 
distribution with equal values of an expectation 
and a variance  determined 
by  a trend  (smooth dependence).   
Using Kolmogorov criterion 
we obtained appropriate agreement 
between distributions of peak values ${\cal P}$ 
at various k
and the exponential distribution. These simulations
confirm that a significance probability 
at certain critical levels 
${\cal P}_{\rm cr}$   
may be estimated as        
%
\begin{equation}
    {\bf P}\{ {\cal P}({\rm k}) > {\cal P}_{\rm cr} \} = 
    \exp \{- {\cal P}_{\rm cr}\},
\label{Pcr}
\end{equation}
%
and Eq.~(\ref{beta})  does
for estimations.
Our simulations have shown 
that the point-like analysis may be treated 
as the limit case  of the standard approach with  
successively narrowed independent bins.
The presence of the trend is a source  
of most powerful long-waved harmonics 
(smallest  k)  in the power spectra
which should be rejected in our consideration. 

The power-spectrum analysis of the ALSs 
is carried out with using Eq.~\ref{P_k}
for successively reduced 
intervals $z_{\rm min} \leq z \leq 4.3$    
($\eta_{\rm min} \leq \eta \leq 1.87$),
the low boundary being shifted  
upward from $z_{\rm min}=0$ to
$z_{\rm min}=2.0$ with a step  
$\delta z_{\rm min}=0.01$.
Filled regions in
the upper panel of Fig.~\ref{Pmax}
indicate positions 
of the main  peaks
with significance  $ \geq 3 \sigma$
on the  $(z_{\rm min}, {\rm k})$-plane
at integer harmonic numbers. 
All marked peaks correspond to the same period
$\Delta \eta = (\eta_{\rm max} - \eta_{min})/{\rm k}$
$= 0.036 \pm 0.002$.   
The lower panel displays  the value  
of peaks P$_{\rm max}$(k)
calculated for all numbers  k
corresponding to the
filled regions on the upper panel.   
Note the presence of periodicity 
in the $\eta$-distribution of ALSs  
for the whole interval of 
$z = 0.0$ -- 4.3\  ($\eta = 0.0$ -- 1.87) 
at the significance level $ \gtrsim 3 \sigma$.
The  highest peaks $(\sim 4.5 \sigma)$
take place
for reduced intervals  $z_{\rm min} \geq 1.64$\
($\eta_{\rm min} \geq 1.148$) 
where statistics is essentially better.    
 
\begin{figure}[t]   
\begin{center}
\epsfxsize=\columnwidth
\epsffile[85 200 515 765]{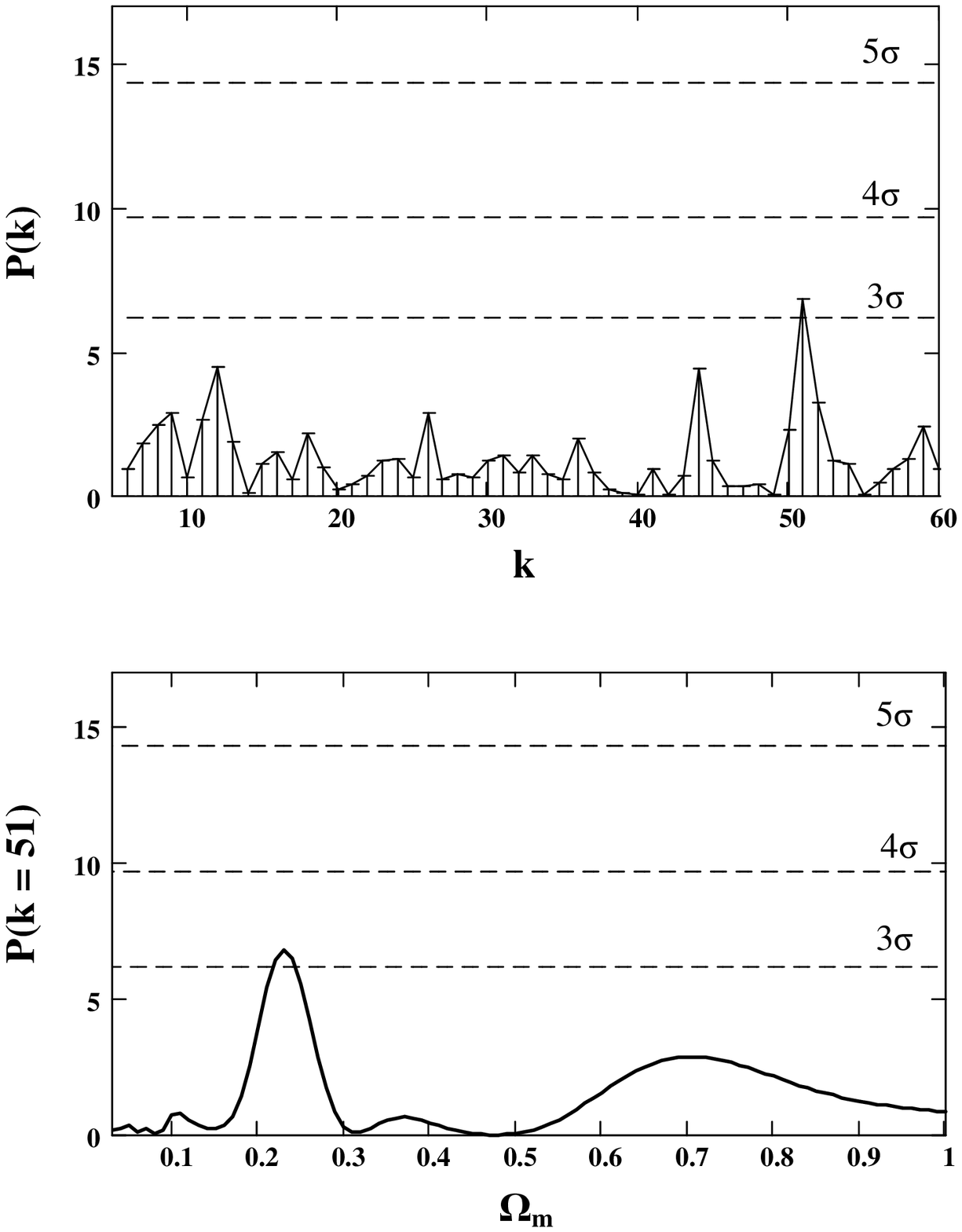}
\caption{  
\footnotesize
{\it Upper panel} represents 
the power spectra  P(k)  
calculated for the whole interval
$\eta = 0.0$ -- 1.87 \, 
($z = 0.0$ -- 4.30) \,
at  $\Omega_{\rm m} = 0.23$;
the main peak corresponds to  k=51
and $\Delta \eta = 0.037$.
{\it Lower panel} plots the dependence  
of the  peak value P(k=51) 
on $\Omega_{\rm m}$. 
}
\label{P-all-k}
\end{center}
\end{figure}
 
\begin{figure}[t]   
\begin{center}
\epsfxsize=\columnwidth
\epsffile[85 200 515 765]{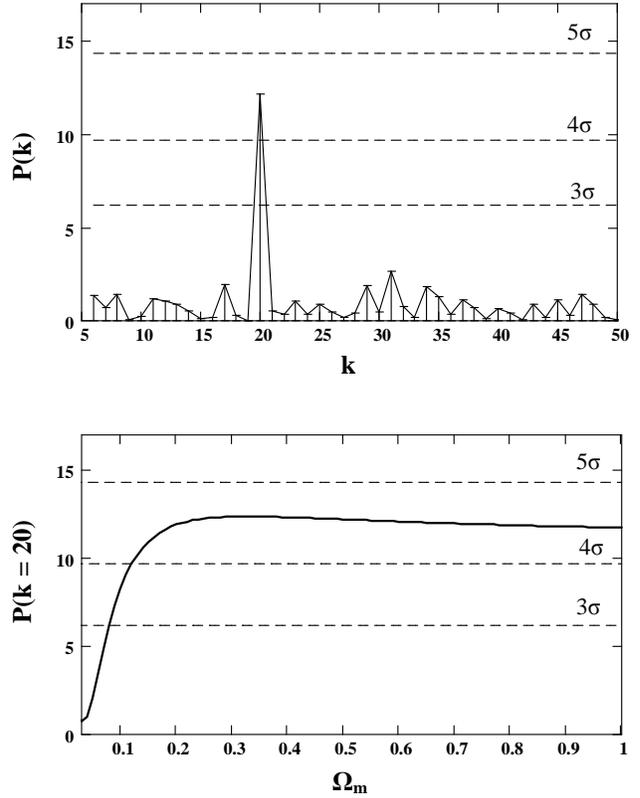}
\caption{
\footnotesize
Same as in Fig.~\protect{\ref{P-all-k}}
but for the most significant peak 
($4.5~\sigma$) in Fig.~\protect{\ref{Pmax}}.
{\it Upper panel}
represents the power spectrum P(k)
calculated  for the interval 
$\eta = 1.148$ -- 1.87 \,
($z = 1.64$ -- 4.30) \, 
at $\Omega_{\rm m} = 0.23$;
the sample contains 
N$_{\rm AS}$ = 1148 ALSs. 
{\it Lower panel} plots the dependence  
of the  main peak  P(k=20) on 
$\Omega_{\rm m}$. 
}
\label{P-part-k}
\end{center}
\end{figure}

For illustration, 
the upper panels in Figs.~\ref{P-all-k} and  
\ref{P-part-k}
demonstrate the power spectra  P(k) calculated 
in a wide region of the harmonic number k
for the whole interval 
$z = 0.0$ -- 4.30\  ($\eta = 0.0$ -- 1.87)  
and for  reduced one 
$z = 1.64$ -- 4.30\  ($\eta = 1.148$ -- 1.87),
respectively.
The lower panel in Fig.~\ref{P-all-k}
displays a relatively sharp dependence   
of the main peak amplitudes P(k=51) 
on the density parameter $\Omega_{\rm m}$, 
which could prevent one to 
bring out  the period $\Delta \eta$ 
in the whole interval of $\eta$, e.g., at  
$\Omega_{\rm m}\sim 0.3$  
(used in Paper~I).     
We assume that this sharp dependence is bounded with
a lack of statistics  at $\eta \la 0.6$, although 
it should be verified in further analysis. 

The peak at  k=20 
($\Delta \eta = 0.036$) 
in the upper panel of Fig.~\ref{P-part-k}
exceeds 
the significance level $4.5 \sigma$.  
The lower panel demonstrates
the dependence of the peak amplitude 
on $\Omega_{\rm m}$. 
One can see that
in contrast with the lower panel 
in Fig.~\ref{P-all-k}
the peak values become 
almost independent of
$\Omega_{\rm m}$ 
at $\Omega_{\rm m} \gtrsim 0.2$.

\section{One-dimensional correlation function}
\label{s:CF}

For verification of the periodicity obtained
in Section~\ref{s:period} 
we calculate additionally a two-point 
correlation function $\xi(\delta \eta)$
for the sample of ALSs. 
In our consideration 
a variable $\delta \eta$  substitutes 
the comoving distance $r$
between pairs of sampled objects
in the standard  two-point 
correlation function $\xi(r)$
(e.g., \citeauthor{dp83} \citeyear{dp83}, 
\citeauthor{r86} \citeyear{r86},  
\citeauthor{mjb92} \citeyear{mjb92}, 
\citeauthor{p93} \citeyear{p93},  
\citeauthor{ls93} \citeyear{p93},  
\citeauthor{ham93} \citeyear{ham93}, 
\citeauthor{kss00} \citeyear{kss00},
and references therein).     
Unlike the standard approach 
the function  $\xi(\delta \eta)$  
is based on the choice of the
single reference center $\eta=0$ ($z=0$)  
associated with the observer. 
Accordingly, all radial (line-of-site) 
points  $\eta_i(z_i)$    
within a layer ($\eta \pm \Delta_\eta/2$),
where $\Delta_{\eta}$ is a width of bin, 
are treated as equivalent ones    
despite of various spatial  
distances between them.    

In this way we count up all pairs 
with fixed relative
line-of-site comoving distance  
$\delta \eta_{i, j} = |\eta_i - \eta_j |$
between  arbitrary ALSs
numerated by $i$ and $j$, which are registered 
in  various  directions.  Thus one can write:
\begin{equation}
     \xi(\delta \eta) = {{\cal N}_{\rm obs}(\delta \eta) \over
                           {\cal N}_{\rm sim}(\delta \eta)} - 1,
\label{xi}
\end{equation}
where ${\cal N}_{\rm obs}(\delta \eta)$
is a number of observed pairs of ALSs  
separated by $\delta \eta = \delta \eta_{i, j}$ 
belonging to the interval $\delta \eta \pm \Delta_{\eta}/2$.
The bin width is chosen as 
$\Delta_{\eta}=0.018$, i.e., 
the same as in low panel of Fig.~\ref{AS91-09}.
${\cal N}_{\rm sim}(\delta \eta)$ 
is a number of crossing pairs between 
the real sample of ALSs 
and the points of a random (Poisson)
sample simulated in the same 
interval of $\eta$  with 
the same smoothed 
distribution (trend) as the real sample.   
The both samples  (real and simulated ones)
have the same number of redshift points.

\begin{figure*}[t] 
\epsfxsize=.8\linewidth
\epsffile[0 220 515 765]{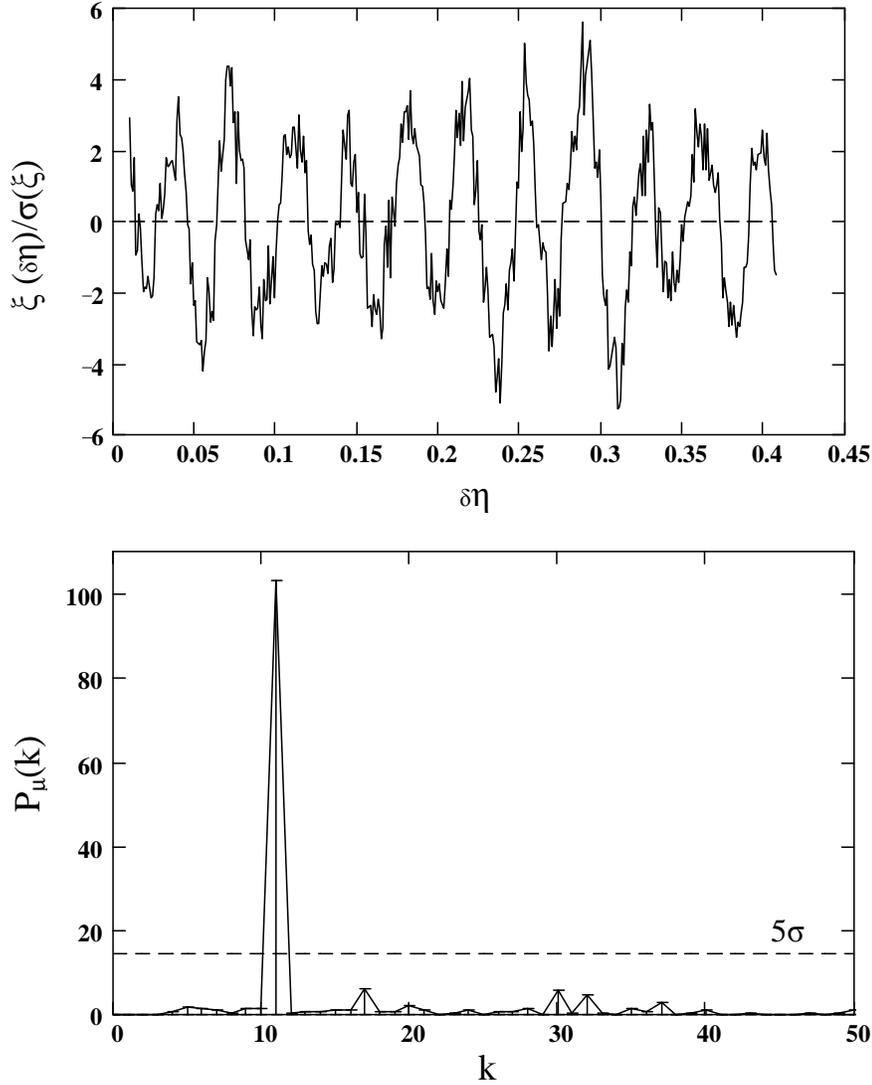}
\caption{
\footnotesize
{\it Upper panel}:
two-points correlation function 
\protect{$\xi (\delta \eta)/\sigma(\xi)$}\,
(see text) versus
the interval of $\delta \eta$  
($0.01 \leq \delta \eta \leq 0.408$)
between the components of pairs 
constituted  from the same sample of 
1148 ALSs  as in 
Fig.~\protect{\ref{P-part-k}}.  
{\it Lower panel}:  
power spectrum  P$_\mu$(k) 
calculated  for 
\protect{$\mu=\mu(\eta)=\xi (\delta \eta)/\sigma(\xi)$}
according to 
Eq.\ \protect{(\ref{Pk_mu})}, 
k is the harmonic number.
} 
\label{XiSig}
\end{figure*}  

Note also that
the correlation function  (\ref{xi})  
differs from the two-point multicentric
correlation  function  which has been calculated for 
both radial  (line-of-sight) and  transverse 
directions by 
\citet{gch09, gms09}  
for the spatial distribution of galaxies
(see also  \citeauthor{kaz10b} \citeyear{kaz10b}).  
On the other hand,
the function (\ref{xi}) differs 
from the correlation functions calculated
by counting  pairs of
objects along separate lines of sight 
(e.g.,  \citeauthor{qby96} \citeyear{qby96}, 
\citeauthor{bj00} \citeyear{bj00}), 
although  the results
of both approaches
seem to be compatible.

The upper panel of Fig.\ \ref{XiSig}
displays the
one-dimensio\-nal (two-point)
correlation function  (\ref{xi})
calculated
in units of the appropriate Poisson error
$\sigma(\xi)=(1+\xi)/\sqrt{{\cal N}_{\rm sim} (\delta \eta)}$    
(e.g., \citealt{pn91}):
\begin{equation}
           {\xi(\delta \eta) \over \sigma(\xi)}
         \approx {{\cal N}_{\rm obs} (\delta \eta) -
            {\cal N}_{\rm sim} (\delta \eta)  \over
        \sqrt{{\cal N}_{\rm sim} (\delta \eta) }}.
\label{xitosig}
\end{equation}

An interval of values 
$\delta \eta$ in  Fig.~\ref{XiSig}\
($0.01 \leq \delta \eta \leq 0.408$) 
is chosen about twice shorter than 
the $\eta$-interval 
($1.148 \leq \eta \leq 1.870$)   
to provide its uniform filling  
by various pairs of ALSs.
One can see the sequence of 
positive and negative peaks
with the significance
$\gtrsim 3 \sigma$ with  respect to zero level.
Let us notice the presence
of a long-range order in the dependence of
$\xi(\delta \eta)/\sigma(\xi)$ on $\delta \eta$.
We discuss this effect in Section~\ref{concl}.  

The lower panel in Fig.\ \ref{XiSig}
represents the power spectrum calculated for 
the  value 
$\mu = \mu (\delta \eta) = \xi(\delta \eta)/\sigma(\xi)$
according to the equation: 
%
\begin{eqnarray}
         {\rm P}_\mu ({\rm k}) & =  & {1 \over N_\mu \hat{U}({\rm k})} \  
                     \left\{ \left[
          \sum_{m=1}^{N_\mu}  \mu_m  \cos \left(
          {2\pi {\rm k} \, \delta \eta_m  \over  \delta \hat{\eta}}
              \right) \right]^2  \right. 
\nonumber                                 \\   
	      & +  & 	      
          \left.   \left[ \sum_{m=1}^{N_\mu}  \mu_m
             \sin \left(
       {2\pi {\rm k} \, \delta \eta_m  \over   \delta  \hat{\eta}}
              \right) \right]^2 \right\},
\label{Pk_mu}
\end{eqnarray}
%
where 
$\mu_m=\mu(\delta \eta_m)$, \,
the values $\delta \eta_m$   
run over a set of points 
$m=1,2,...N_\mu$  of the 
variable $\delta \eta$ from 0.01 to 0.408. 
Here we employ 
the sliding bins $\Delta_\eta=0.018$ 
with  centers at $\delta \eta_m$  which
transit from one center to another one by successive shifts 
$h_{\eta} = 0.001$;
$N_\mu= 399$ is the number 
of the points $\delta \eta_m$ within the whole  
interval $\delta \hat{\eta} = 0.398$. 
The value $\hat{U}$(k) in the denominator of (\ref{Pk_mu})
may be  presented as
\begin{equation}
         \hat{U}({\rm k}) = \left\{ \begin{array}{c l l}
	              U({\rm k})  & {\rm at} & U({\rm k}) > 1  \\
	               1    & {\rm at} & U({\rm k}) \leq 1,
		       \end{array}
		        \right. 
\label{hatU}
\end{equation}
where
\begin{equation} 
       U({\rm k}) = \left[   \sin ( { \pi {\rm k} \Delta_{\eta} \over \delta \hat{\eta}})
			\,  {\delta \hat{\eta}  \over  \pi {\rm k} \Delta_{\eta}} 
			 \right]^2  { \Delta_{\eta} \over h_{\eta}}.  
\label{Uk}
\end{equation}
It is known from spectral analysis 
(e.g., \citeauthor{bath74} \citeyear{bath74}),
that the  sliding-average  procedure
corresponds to a low-frequency filtration 
with a rectangular function as a filter. 
Fourier transform 
of this function
is represented by 
$\sin (\omega/2) / (\omega/2)$,
where $\omega$ is a  dimensionless Fourier frequency.
Note that in the considered case     
a factor $\Delta_{\eta} / h_{\eta}$
in Eq.~(\ref{Uk}) is equal to $18$.  
Our special simulations 
of random catalogs of points
have shown that the normalization       
factor (\ref{hatU}),\ (\ref{Uk}) in the denominator 
of  expressions  analogous to Eq.~(\ref{Pk_mu}) 
compensates a low-frequency distortion 
through
the sliding-average filtration and
yields  the exponential   
probability functions given by Eq.~(\ref{Pcr}).
Therefore we use  Eq.~(\ref{beta}) 
for estimations of the confidence probability
of peaks in the power spectrum. 

\begin{figure}[t]   
\begin{center}
\epsfxsize=\columnwidth
\epsffile[85 200 515 765]{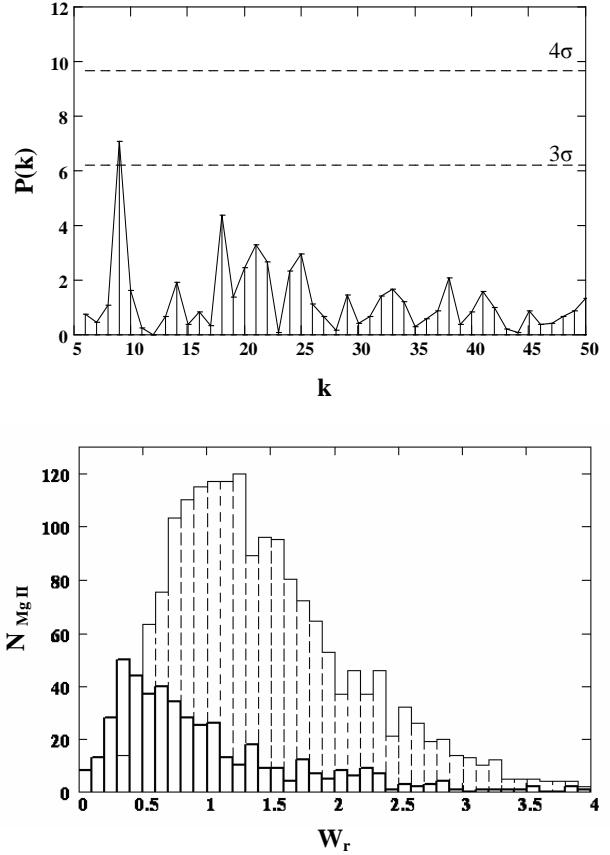}
\vspace{1.0cm}
\caption{
\footnotesize
{\it Upper panel}:  
Power spectrum  P(k)  
($5 \leq {\rm k} \leq 50$) 
calculated in the point-like statistical approach 
(Eq.~\protect{(\ref{P_k})}) with selection of
961 absorption systems Mg~II from the catalog  
by \cite{bouche06} within the interval 
$0.38 \leq z \leq 0.8$\ 
($0.35 \leq \eta \leq 0.67$) 
satisfying a condition of 
W$_{\rm r} \leq 1.4$~\AA.   
{\it Lower panel}:
Comparison of two  histograms 
displaying  
distributions of 478  Mg~II  absorption systems from 
the catalog by \cite{rkv03} (thick solid lines)  
and 1806 Mg~II systems from the catalog 
by \cite{bouche06} (long dashed lines)
with respect to  
the rest-frame equivalent width ${\rm W}_{\rm r}$.
} 
\label{MgII}
\end{center}
\end{figure}   
\begin{figure}[t]    
\begin{center}
\epsfxsize=\columnwidth
\epsffile[85 200 515 765]{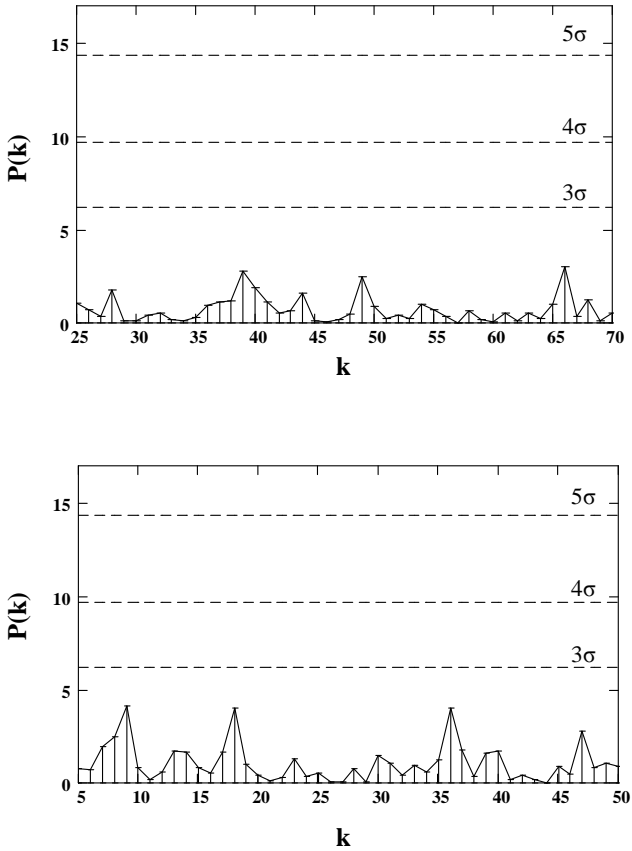}
\vspace{0.5cm}
\caption{
\footnotesize
{\it Upper panel}:  
Power spectrum P(k)  
calculated  with  using  
Eq.~\protect{(\ref{P_k})} for 730 QSOs 
($0.035 \leq  z_e \leq 5.01$) 
from the catalogue \protect{\cite{rkv03}},
whose  spectra
include 2322 ALSs  used in the present analysis.       
An  interval  of harmonic numbers    
$25 \leq {\rm k} \leq 70$ 
corresponds to comoving  scales
$(233.3 - 83.3)~h^{-1}$~Mpc,  
at $\Omega_{\rm m}=0.23$.  
{\it Lower panel}:  
Same as the {\it upper panel} but
for 961 QSOs  
($0.42 \leq  z_e \leq  2.90$) 
from the catalogue 
~\protect{\cite{bouche06}},
whose spectra  include  
961 systems of  Mg~II
used for the calculations
of  P(k) in 
Fig.~\protect{\ref{MgII}}.
An interval of  harmonic numbers    
$5 \leq {\rm k} \leq 50$ 
corresponds to  comoving  scales
$(712.9 - 71.3)~h^{-1}$~Mpc.  
} 
\label{QSOs}
\end{center}
\end{figure}  
    
In result we obtain a strong  peak of 
P$_\mu$(k) at  k = 11
on the significance level 
well exceeding  $5 \sigma$.
This peak  manifests 
the periodicity 
of the correlation function
with the period
$\Delta \eta = 0.036 \pm 0.002$ which is in
agreement
with the periodicity
discussed in Sect.\ \ref{s:period}.

\section{Periodicity in distribution of Mg II systems}
\label{s:MgII}

It should be noted in this context that 
a set of statistical investigations of resonance
absorption doublets Mg~II ($\lambda$ 2796, 2803 \AA) on the base of SDSS 
catalog have been performed
recently by a few groups
(e.g., \citeauthor{bouche06} \citeyear{bouche06},
\citeauthor{prprb06} \citeyear{prprb06},
\citeauthor{menard08} \citeyear{menard08},
\citeauthor{lundetal09}  \citeyear{lundetal09}).
It follows from their results that 
the spatial distribution  
of strong doublet lines Mg~II with 
W$_{\rm r}^{\lambda2796} > 1$~\AA \   
differs from the distribution of relatively weak lines 
W$_{\rm r}^{\lambda2796} < 1$~\AA, where W$_{\rm r}^{\lambda2796}$ 
is the rest frame equivalent width of the line $\lambda=2796$~\AA .   
It seems to be plausible 
that the strong and weak Mg~II absorbers 
are associated with different galaxy populations
which display different evolving 
(e.g.,  \citeauthor{lundetal09}  \citeyear{lundetal09})
and morphological 
(e.g., \citeauthor{kacp07}  \citeyear{kacp07, kacp08})
properties.  Therefore one could anticipate 
that radial distributions of both 
types of absorption systems  correspond to different         
tendencies of clustering and different periodicities.   
For instance, weaker absorbers appear to be  
spreaded over space-time more homogeneously
and with weaker effects of galaxies evolution.   

Let us note that the main result 
of the present paper
is bound up with  relatively weak ALSs. 
It is illustrated  on the lower panel of Fig.~\ref{MgII} 
by the histograms displaying  the  distribution 
of  478 Mg~II absorption doublets $(\lambda2796, 2803)$  
involved in our analysis  
with respect to  
W$_{\rm r}^{\lambda2796}$
and the analogous distribution of 1806 Mg~II  
systems from the catalog 
by \cite{bouche06} 
based on the third data release (DR3) of the SDSS. 
The  distribution of Mg~II systems 
from our catalog  has a maximum at
W$_{\rm r}^{\lambda2796} \lesssim 0.4$~\AA\ 
while the distribution  
of the  Mg~II  absorbers 
from the catalog of \cite{bouche06}
shows a maximum at  
W$_{\rm r}^{\lambda2796} \gtrsim 1.0$~\AA, 
i.e., the  former  one 
is noticeably shifted towards  
lower values of W$_{\rm r}^{\lambda2796}$.
Let us note that the more extended catalogue 
of Mg~II systems by \cite{prprb06}
also based on SDSS data 
accumulates still stronger lines with 
W$_{\rm r}^{\lambda2796} \geq 1$~\AA\ 
which are still less consistent
with our sample of ALSs.
As discussed above it is likely that 
these two samples   
correspond  predominantly to  
different populations of 
galaxies.  
    
Nevertheless, we estimate here the presence 
of the same periodicity 
in the $\eta$-distribution 
of Mg~II systems as discussed 
in Sections~\ref{s:period} and  \ref{s:CF}  
restricting ourselves by 
relatively weak systems 
(W$_{\rm r}^{\lambda2796} \lesssim 1$~\AA).
The upper panel in Fig.~\ref{MgII} 
represents the power    
spectrum  ${\rm P}(k)$ calculated  
according to the point-like approach of Eq.~(\ref{P_k}). 
We employ  N$_{\rm tot}$=961 systems 
from the catalog  
of \cite{bouche06} at 
W$_{\rm r}^{\lambda2796} \leq 1.4$~\AA\
within the interval 
$0.35 \leq \eta \leq 0.67$  $(0.38 \leq z \leq 0.80)$.     
One can see the peak P$(k)$ at $k=9$ 
for the same period
$\Delta \eta = (\eta_{\rm max} - \eta_{min})/k$
$= 0.036 \pm 0.002$  as in  Figs.~\ref{Pmax}--\ref{XiSig}. 

Additionally,  we test possible selection effects which might 
induce  periodicities in the initial
samples of QSOs.  Examples of such simulated periodicities
are given, e.g.,  by  \cite{hart09} and \cite{bc10}.
The upper panel in Fig.~\ref{QSOs}  represents 
the power spectrum P(k)   
calculated according to  
Eq.~(\ref{P_k}) for 730 original QSOs  
within an interval of 
$\eta_e = 0.035 - 1.98$\ 
$(z_e = 0.035 - 5.01)$ 
sampled for the  
statistical treatment of 
2322 ALSs. 
We see that in the chosen range of harmonic numbers,
k=25 -- 70, there are no significant peaks in the
power spectrum of the  QSOs. Note that 
the period   $\Delta \eta = 0.036$ 
indicated in Sections~\ref{s:period} and  ~\ref{s:CF}
corresponds to the harmonic number k=54.   
The lower panel demonstrates similar power spectrum
calculated for 961 QSOs    
within an interval of  
$\eta_e = 0.39 - 1.58$\   
$(z_e = 0.42 - 2.90)$  
from the catalog by  \cite{bouche06}.
We use their spectra 
for the test of periodicity of
961 relatively weak (W$_{\rm r} \leq 1.4$~\AA\ ) 
systems Mg~II. The power spectrum also does not    
contain significant peaks. 
Note  that  the period   $\Delta \eta = 0.036$ 
corresponds to the harmonic number k=33.  
On the other hand, possible simulated periodicities  
at $\Delta z_e = 0.258, 0.312$, and 0.44 
in $z_e$-distribution of QSOs   
indicated  by \cite{hart09} for SDSS data  
correspond to harmonic numbers k$\approx 10$, 8, and 6, respectively. 
Additional analysis of possible selection effects
was  carried out in Paper~I.

\section{Conclusions and discussion}
\label{concl}

On the base of the updated
catalog of absorption systems \citep{rkv03}
we have performed the 
statistical analysis  of 2322
absorption-line systems (ALSs)  
detected in the  QSO spectra 
in the redshift range $z = 0.0-4.3$.
The ALS  redshifts  are 
averaged within the velocity 
interval $\delta v = 500$~km~s$^{-1}$ 	
along  each line of sight.
This velocity interval  
containing one averaged point 
corresponds to the  
spatial scales $\sim 5-6~h^{-1}$~Mpc  
(at $z=0.0$--4.3)
of  comoving distances 
for  $\Lambda$CDM model 
at  $\Omega_{\rm m}=0.23$. 
The  averaged points  
can trace  possible   
large-scale  
variations of absorbing matter.

\noindent
The main results can be summarized 
as follows: \\  
(1)  The distribution of the ALSs 
relative to the dimensionless comoving  
distances $\eta(z)$   
given by Eq.~(\ref{eta}) displays    
a statistically significant pattern 
of alternating maxima (peaks) and minima
(dips) against a smoother  curve (trend).   
This sequence of peaks and dips 
comprises a certain periodical component 
with a period of $\Delta \eta = 0.036 \pm 0.002$.
The power spectrum calculated
according to Eq.\ (\ref{P_k}) for
the whole interval of $\eta$ 
($\eta = 0.0$ -- 1.87)   
contains the peak at the significance level
exceeding $3 \sigma$ 
relatively to the hypothesis 
of the Poisson $\eta$-distribution. 
In the most statistically  
representative interval
$\eta = 1.148$ -- 1.870 
($z=1.64$ -- 4.3)
the same periodicity
reveals itself at the significance 
level $4.5 \sigma$. 
Still more prominent peak corresponding to 
the same period $\Delta \eta$ arises
in the power spectrum  calculated 
for the two-point  correlation function
$\xi (\delta \eta)$   with use of Eqs.\ 
(\ref{Pk_mu} -- \ref{Uk}). 

\renewcommand{\arraystretch}{1.2}
\begin{table*}[t] 
\caption[]{
Characteristic scales of LSS relatively 
to comoving system in $\Lambda$CDM-model} 
\label{tab:scale}
\begin{center}
\begin{tabular}{ c c c c c c }
\hline
\hline
Objects & Surveys &  Redshifts  &  $\Lambda$CDM  & Scales  &  References  \\ 
  \,  & Catalogs  & \,  &  $\Omega_{\rm m}$     &  $h^{-1}$~Mpc  &   \,  \\ 
\hline
92,952~G~$^{1)}$ & SDSS~DR1~$^{5)}$ (radial)  &  $z < 0.14$  &  \,  & 
74~${\pm  17}$ & \, \\
 \,      &   \,    &   \,    &  0.3  &  \,   &  ~~$^{14)}$  \\ 
16,756~G  & LCRS~$^{6)}$  &  $z < 0.11$ &   \,  & 60~${\pm 10}$ &  \,   \\
\hline
1324~CG~$^{2)}$  & Abell~CG  &  \, & \, & \,  &  \,  \\
   +    &   +    &   $z < 0.28$   & ~~$^{11)}$  & 115~${\pm 7}$   &  ~~$^{15)}$  \\
284~X-ray~CG~$^{3)}$ & RBS~$^{7)}$  &  \,  &  \,  &  \,	&  \, \\  
\hline
229,193~G  &  2dF~GRS~$^{8)}$   &   \, &  \,  & 73.4~$\pm 5.8$   &  \,  \\
   +   &   +  &  $z < 0.3$  &  ~~$^{12)}$   &  +   &   ~~$^{16)}$   \\   
427,512~G  &   SDSS~DR5   &  \,   &  \, &  127~$\pm 21$   &  \,  \\
\hline
143,368~G  &  2dF~GRS  &    $z=0.2$  &  \,  &  112.9~$\pm 3.3$  &  \,   \\
   +   &    +     & \,    &  0.23~$^{13)}$   &  \,  &   ~~$^{17)}$  \\
77,801~LRG~$^{4)}$  &   SDSS~DR5      &   $z=0.35$ &  \,  & 104.7~$\pm 3.2$ &  \,   \\                
\hline
    \,   &   \,   &    $0.15 \leq z \leq 0.30$  &  \,  & 110.3~$\pm 3.9$  &  \,  \\  
75,000~LRG  &  SDSS~DR6    &  \,  & 0.25  &  \,   &    ~~$^{18)}$  \\
   \,    &    \,  &   $0.40 \leq z \leq 0.47$ & \,  & 108.9~$\pm 4.0$ &  \,   \\
\hline	
105,831~LRG &  SDSS~DR7 & $0.16 <  z  < 0.36$ &  0.25 &  $101.7 \pm 3.0$ &
 ~~$^{19)}$  \\
\hline
2167~ALS &  RKV~$^{9)}$ & $ 0 < z \leq 4.3$ &  0.23 & $108 \pm 6$ &  present \\
961~Mg~II &  Bouch\'{e}~$^{10)}$ & $0.38 \leq z \leq 0.80$ &  0.23  & $108 \pm 6$ &
 work  \\  
\hline
\end{tabular}
\end{center}
\vspace{0.5cm}
\begin{flushleft}
\     \\
{\footnotesize
$^{1)}$   ~G  --  Galaxies; 
$^{2)}$   ~CG -- clusters of galaxies;    
$^{3)}$   ~X-ray CG -- X-ray clusters of galaxies;  
$^{4)}$   ~LRG -- Luminous red galaxies;    
$^{5)}$   ~SDSS -- Sloan Digital Sky Survey;  
$^{6)}$   ~LCRS -- Las Campanas Redshift Survey; 
$^{7)}$   ~RBS  -- ROSAT Bright Survey; 
$^{8)}$   ~2dF GRS -- 2dF Galaxy Redshift Survey;
$^{9)}$   ~RKV -- \citet{rkv03};  
$^{10)}$  ~\citet{bouche06};  
$^{11)}$  ~distances between clusters were calculated using the \citet{m58} 
formula with the deceleration parameter $q_0 = 0.5$ \citep{einastetal01}; 
$^{12)}$   ~distance intervals were calculated using $\Delta r = c~\Delta z/H_0$;     
$^{13)}$   ~the scales $r_{\rm s}$ have been recalculated  for this 
Table at  $\Omega_{\rm m}=0.23$;
~~$^{14)}$   ~\citet{doretal04};    
~~$^{15)}$   ~\citet{tagoetal02};
~~$^{16)}$   ~\cite{harhir08};
~~$^{17)}$   ~\cite{percetal07a};
~~$^{18)}$   ~\cite{gch09};   
~~$^{19)}$   ~\cite{kaz10a}.   
Sign $+$ means that the data of both catalogs 
and both types of objects
were used, as well as  both scales were revealed for the same sample 
of data. \\  
}
\end{flushleft}
\end{table*}
 
(2) The same period has been revealed also 
in the power spectra of 961~Mg~II\  $(\lambda2796, 2803)$ 
absorption systems     
from the catalog  by \cite{bouche06} 
within the interval 
$0.35 \leq \eta \leq 0.67$  $(0.38 \leq z \leq 0.80)$  
calculated for  
relatively small
rest frame equivalent widths 
W$_{\rm r}^{\lambda2796} \leq 1.4$~\AA.
Although the appropriate peak in the power spectra
is not very significant (slightly exceeds $3 \sigma$)
this is comparable with the peak obtained 
in Section~\ref{s:period}
for the sample of ALSs   
within the whole interval   
$\eta = 0.0$ -- 1.87 ($z = 0.0$ -- 4.3).      
It may  evidence for availability 
at least a weak  periodical component 
at  $z < 1$. 

(3)  The special feature of the results  
is the appearance of the long-range 
(or intermediate-range) 
order visually  displayed by the two-point correlation
function in Fig.~\ref{XiSig}. 
This order corresponds to
some stability of the phase of
periodical component which leads to 
an invariance 
of the correlation function under 
translations   
by a distance multiple 
of the period  $\Delta \eta$.   
The significant peak at the lower panel 
in Fig.~\ref{XiSig} may be treated as  
a  measure of  such an  order.

(4) The dimensionless period $\Delta \eta$ 
obtained here corresponds to the
spatial scale of line-of-sight comoving distance 
$D_{\rm c} = c/H_0 \times \Delta \eta =$ 
$108 \pm 6~h^{-1}$~Mpc.   
According to Table~\ref{tab:scale} this
scale $D_{\rm c}$ is consistent
with a  characteristic scale
of the Large Scale Structure (LSS) in the spatial 
distribution of galaxies and clusters of galaxies
at relatively small redshifts $z <0.5$. 
On the other hand, the appropriate  
temporal interval
$T_{\rm c} = 1/H \times \Delta \eta =$
$350 \pm 20~h^{-1}$~Myr
is consistent with recent  
results of \citet{ak08}. 

Summarized results admit  both 
interpretations of the 
periodicity:  spatial one -- 
appearance 
of  partly ordered spatial structures 
of matter in the early Universe (see below)
or  temporal one -- generation of  
some temporal wave process
in the course of the 
cosmological evolution. 
For instance, \citet{krv00} and Paper~I 
argued rather in favour of
the temporal interpretation of the features of 
ALS distribution. In this paper, however, 
we follow a spatial consideration.
Note that both hypotheses 
have to obey to the cosmological 
principle (e.g, \citealt{p93}), 
i.e., partly  ordered structures or 
periodical temporal process could be observed
in any spatial  
points  of the Universe.

A special wave of interest  
to the effects of  
periodicity in the redshift distribution 
of galaxies  was initiated  by \citet{beks90}.    
Their pencil-beam surveys
near the Galactic poles displayed  a
periodicity on a scale about $130~h^{-1}$~Mpc 
which might be interpreted 
either as a pure spatial  quasi-periodic 
pattern of LSS  constituents 
(set of clumps or walls and voids;
e.g., \citeauthor{ksmf90} \citeyear{ksmf90},  
\citeauthor{kp91} \citeyear{kp91}, 
\citeauthor{vdw91} \citeyear{vdw91},  
\citeauthor{dbps92} \citeyear{dbps92}, 
\citeauthor{yosh01}  \citeyear{yosh01}  
and references therein)
or as a spatial-temporal sequence of  
pronounced and depressed epochs
which could become apparent 
in the distribution of matter
(e.g.,  \citeauthor{m91} \citeyear{m91}, 
\citeauthor{hkk08}  \citeyear{hkk08}).
 
Table~\ref{tab:scale} represents the examples of   
characteristic scales 
obtained in last years by a few groups of authors
employing  statistical analysis 
of galaxies and clusters of galaxies. 
The characteristic scales in the last but one column 
may be approximately 
subdivided  into two groups: 
the scales belonging to 
the interval $70 \pm 20~h^{-1}$~Mpc and 
the scales within $120 \pm 20~h^{-1}$~Mpc.     
The period $D_{\rm c} = 108 \pm 6~h^{-1}$~Mpc    
gets into the latter group.
Especially good consent occurs 
with results obtained by  
\citet{percetal07a}, \citet{gch09} and \citet{kaz10a} 
for the characteristic scale 
of the Baryon Acoustic Oscillations (BAOs)
in the large-scale distribution of matter
(see also \citeauthor{martetal09} \citeyear{martetal09}).

Actually, the BAOs are discussed in literature 
in connection with a scale of the sound 
horizon at the epoch of recombination   
(e.g., \citeauthor{bg03} \citeyear{bg03}, 
\citealt{percetal07b}) 
which may be measured as
a single peculiarity (bump) in the spatial correlation 
function (e.g., \citeauthor{eisen05}  \citeyear{eisen05}).  
On the other hand, this spatial scale  
manifests itself  as a series of regular variations
imprinted on the power spectrum
of the distribution of galaxies 
(e.g., \citeauthor{eih98}  \citeyear{eih98},  
\citeauthor{eiht98}  \citeyear{eiht98}, 
\citeauthor{percetal07b} \citeyear{percetal07b}).
By contrast, we discuss here 
the  regular 
spatial variations providing one prominent peak 
in the power spectra.     
In this context let us  note  that 
the correlation function 
calculated  by  \citet{martetal09}
shows a hint of a possible secondary peak 
at a scale of
$\sim 170~h^{-1}$~Mpc. 
Being confirmed 
such a tendency would be in consent
with the hypothesis that at least 
an intermediate-range 
order may be present 
in the spatial distribution of ALSs. 
 
Still more pronounced regular 
spatial variations 
of the standard (see Section~\ref{s:CF})
two-point correlation function 
of galaxy superclusters were obtained, e.g., 
by \citet{einetal97a} and \citet{tagoetal02}. 
These authors indicated the possibility 
of  a long-range (or intermediate-range) 
order with a characteristic  period  of 
$115 \pm 7~h^{-1}$~Mpc 
in the spatial distribution of 
superclusters: 
the largest relatively isolated bounded systems.
Their results are qualitatively consistent 
with the two-point 
correlation function shown in Fig.~\ref{XiSig}.
\citet{einetal97b} considered 
a model  of a quasi-regular lattice
with random distribution 
of clusters and superclusters
along lattice edges. They showed that  
the spatial correlation function
for a partly regular structure  
displays oscillations similar
to those  revealed  in the  
correlation function of superclusters.

By analogy with \citet{einetal97b} 
we consider in the Appendix   
two toy models of partly regular
structures with essential random
components  and  calculate the power spectrum 
of radial distribution of points.  
The power spectra
for both models 
display the peaks  with 
approximately the same 
significance
as in Fig.~\ref{P-part-k}.
It illustrates that rather
moderate spatial ordering  
might  yield  the  
periodicity of radial distribution 
similar to the sample
of ALSs  discussed here.   

It should be emphasized that we represent
only visual examples of  partly ordered structure
without strict arguments in favour of the
spatial interpretation of the periodicity obtained. 
This  interpretation  should  be  verified 
by special investigations in a future.

\textit{Acknowledgments}
We thank A.Y.\ Potekhin for technical assistance
in preparation of the paper. 
The work has been supported partly
by the RFBR (grant No.\  08-02-01246a), 
and  by the State Program "Leading Scientific Schools of RF" 
(grant NSh\ 3769.2010.2).
 


\begin{appendix}
\twocolumn
\section{Toy models for partly ordered SC lattice}

\begin{figure}[t]   
\vspace{0.5cm}
\begin{center}
\epsfxsize=\columnwidth
\epsffile[85 150 515 775]{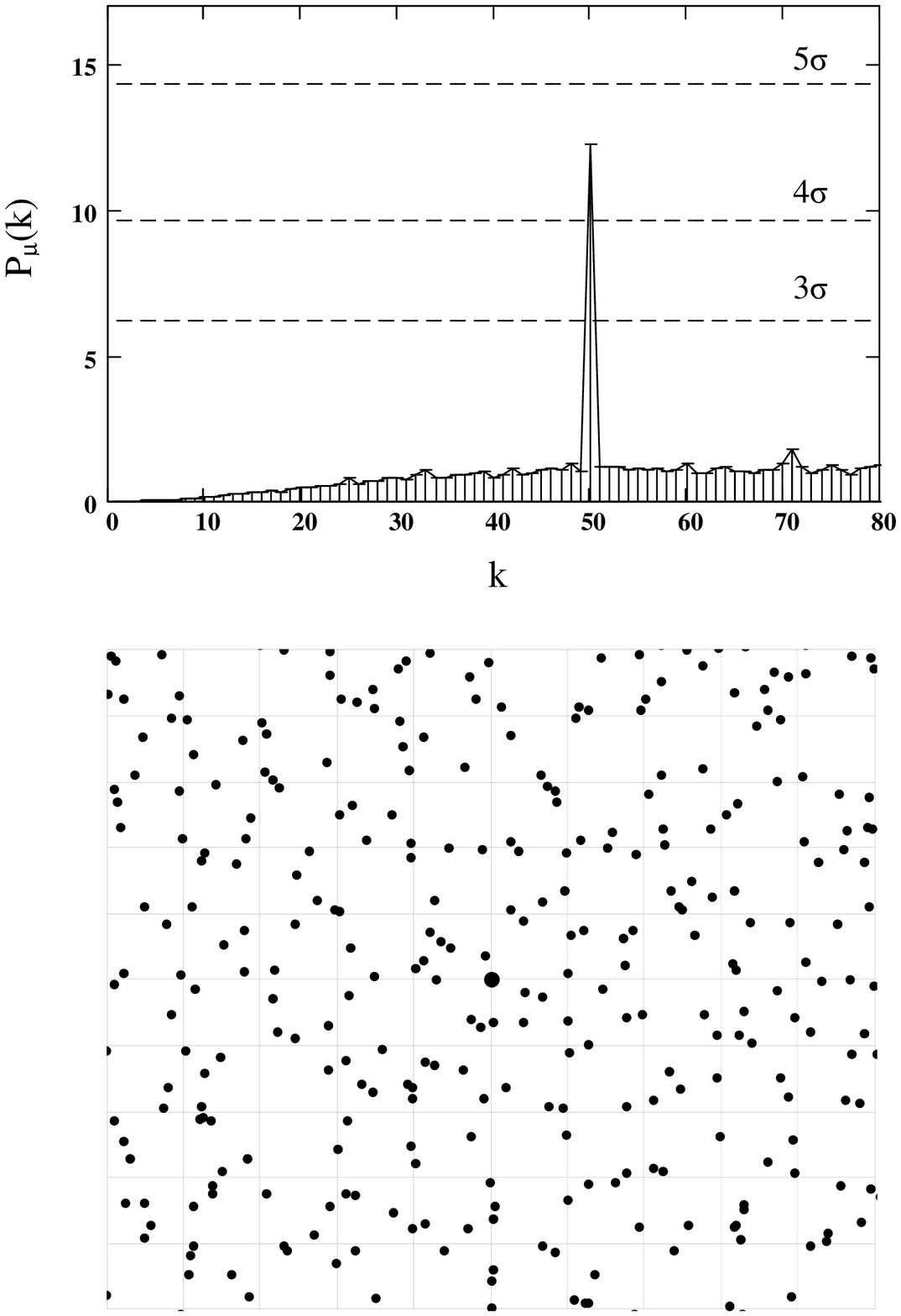}
\vspace{.0cm}
\caption{
\footnotesize
{\it Upper panel}:
Mean value of 100 realizations of 
power spectra  calculated for the radial
distribution function Eq.~(\ref{mu_R}) 
in the  case of a {\it cloud-like} lattice (see text) 
constructed as superposition of 3 random 
cubic lattices with the standard 
deviation $\sigma_{\rm p}=0.3$
from simple cubic.         
{\it Lower panel}:  fragment of
2-dimensional illustration displayed 
one of 100 realizations 
of the cloud-like lattice
(calculated for the upper panel);
points represent lattice vertices
and the filled circle indicates 
the center of the sphere.  
} 
\label{cloud}
\end{center}
\end{figure}
\begin{figure}[t]    
\vspace{.5cm}
\begin{center}
\epsfxsize=\columnwidth
\epsffile[85 150 515 775]{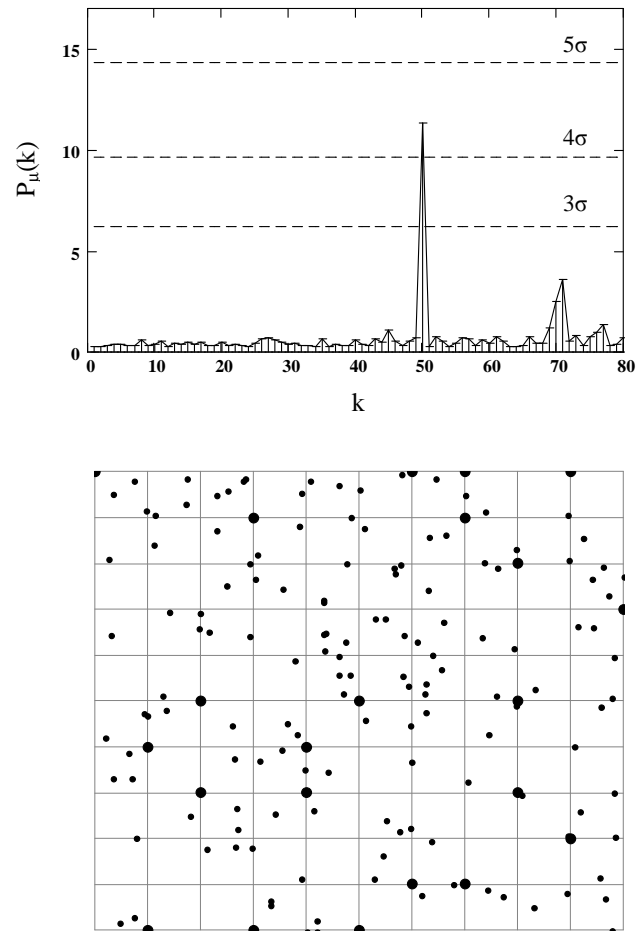}
\vspace{.5cm}
\caption{
\footnotesize
Same as in Fig.~\protect{\ref{cloud}}
but for superposition of a 
{\it hole-like} lattice  
with a parameter $u=0.2$ (see text) and the 
field of random Poisson points 
with  sixfold ($e=6$) excess  
of random points over 
{\it hole-like} lattice vertices.     
{\it Upper panel}: mean value of 100 realization
of power spectra; 
{\it lower panel}: filled circles stand for 
the vertices of the hole-like lattice.  
} 
\label{hole}
\end{center}
\end{figure}

Figs.~\ref{cloud} and \ref{hole} represent two models
of simulations performed for partly ordered structure of points
based on the simple-cubic (SC) lattice
and the results of appropriate  
calculations of power spectra.    

Fig.~\ref{cloud} displays a {\it cloud-like} distribution
of points around the vertices of the  SC lattice
with unit lattice constant.
We consider  a superposition of 3 similar lattices
with coordinates of the vertices represented as
$x_v = i_v + \delta x_v,\,  y_v = j_v + \delta y_v,\, 
z_v = k_v + \delta z_v$, where $i_v,\, j_v$, and $k_v$ are
integer numbers within the interval from $-50$ to $+50$
with respect to the zero point and 
$\delta x_v$, $\delta y_v$, and $\delta z_v$ are random    
numbers with the Gaussian distribution and 
standard deviation $\sigma_v=0.3$. In other words
we have a cloud of three random vertices instead of  
one lattice vertex.      

The radial distribution function 
for the random points with radial coordinate
$r_v=\sqrt{x_v^2+y_v^2+z_v^2}$
within a sphere restricted by the radius
$R=50$ can be calculated as a value 
\begin{equation}
            \mu=\mu_{\rm R}(r_l) =
         {N_{\rm V}(r_l) - n_0\, 4 \pi r_l^2 \Delta r_l  
	 \over  \sqrt{n_0\, 4 \pi r_l^2 \Delta r_l}},
\label{mu_R}
\end{equation}
where $r_l$ is a mean radius of $l-$th concentric layer
with fixed radial width (bin) $\Delta r_l = 0.1$, 
$N_{\rm V}(r_l)$ is a number of vertices within
a set of non-overlapping layers, $n_0$ is a mean density
(number per unit volume) of vertices.   

We calculate the power spectra 
for each realization of $\mu$  given by (\ref{mu_R})
using a version of Eq.~(\ref{Pk_mu}). 
In this version
$r_{\rm m}$  substitutes for $\delta \eta_{\rm m}$ 
and the radius $R$
stands for $\delta \hat{\eta}\, $; 
$\mu_{\rm m}=\mu_{\rm R}(r_{\rm m})$,
where  the value of $r_{\rm m}$ runs over
a uniform set of points  
(with a step 0.10)
$r_{\rm m}=0.05, 0.15, 0.25,...49.95$  
numerated by $m=1,2,....N_\mu$  and  
$N_\mu=500$. By contrast with 
Eq.~(\ref{Pk_mu}) we  apply here the approach 
of independent bins (bin=0.1) and  put  
$\hat{U(k)} = 1$ in the denominator.

The upper panel in Fig.~\ref{cloud}
displays the mean  value
of the power spectra calculated for
100 realizations of the radial distribution 
function (\ref{mu_R}).  The peak with 
significance  $\sim 4.5 \sigma$  
(cf.  Fig.~\ref{P-part-k})  indicates  
the presence 
of oscillations with a period 
equal to unity (lattice constant).         

Similar calculations are
represented in Fig.~\ref{hole} 
where we use so named 
{\it hole-like} lattices 
constructed on the base of SC 
lattice $x_v = i_v,\,  y_v = j_v,\, 
z_v = k_v $,  where $i_v,\, j_v$, and $k_v$ are
introduced above 
with a random removing of some vertices.
Parameter $u=0.2$ is a ratio of the numbers
of residual to initial lattice vertices.  
Additionally we immerse
the hole-like lattice
into the field of random Poisson points
with sixfold  excess of the points
over the number of residual vertices ($e=6$). 

Using the version of   
Eq.~(\ref{Pk_mu}) 
discussed above,
we calculate 
the power spectra for different realizations
of the combined 3-dimensional  
distribution of points. The 
mean value of the power spectra  calculated 
for 100 realizations of the radial 
distribution function (\ref{mu_R})
is represented in the upper panel of 
Fig.~\ref{hole}. One can see  
the peak with 
significance  $\sim 4.5 \sigma$ 
similar to the case of cloud-like 
lattice. 

The lower panels in Figs.~\ref{cloud} and \ref{hole}
represent 2-dimensional illustrations 
of the partly ordered structures used in the 
power-spectrum calculations. Let us emphasize 
a  low  degree of visual ordering of
the structures discussed.

\end{appendix}


\begin{thebibliography}{99}
 
\bibitem[Aref'eva \&  Koshelev(2008)]{ak08}
Aref'eva\ I.~Ya., \& Koshelev A.~S., 2008, JHEP, 9, 68 (arXiv:0804.3570)

\bibitem[Bath(1974)]{bath74}
Bath\ M.,  1974, Spectral analysis in geophysics, Elsevier Science


\bibitem[Bell \& Comeau(2010)]{bc10}
Bell\ M.~B., \& Comeau\ S.~P., 2010,  Ap\&SS, 326, 11
    

\bibitem[Blake \& Glazebrook(2003)]{bg03}
Blake\ C., \& Glazebrook\ K., 2003, \apj, 594, 665
 
   
\bibitem[Broadhurst et al.(1990)]{beks90}
Broadhurst\ T.~J., Ellis\ R.~S., Koo\ D.~C., \& Szalay\ A.~S., 1990,
Nature,\ 343, 726 

    
\bibitem[Broadhurst \& Jaffe(2000)]{bj00}
Broadhurst\ T., \& Jaffe\ A.~H.,  2000,
in  Mazure\ A.,  Le~F\'evre\ O.,  Le~Brun\ V., eds,
Clustering at High Redshift,
ASP Conference Series,  200,
p. 241 

 
\bibitem[Bouch\'{e} et al.(2006)]{bouche06}
Bouch\'{e}\ N., Murphy\ M.~T., P\'{e}roux\ C., Csabai\ I., \&
Wild V., 2006, \mnras, 371, 495


\bibitem[Davis \& Peebles(1983)]{dp83}
Davis\ M., \& Peebles\ P.~J.~E., 1983,\   \apj,\ 267, 465

     
\bibitem[Dekel et al.(1992)]{dbps92}
Dekel\ A., Blumenthal\ G.~R., Primack\ J.~R., \& Stanhill\ D.,  1992,
\mnras,\ 257, 715 


\bibitem[Doroshkevich et al.(2004)]{doretal04}
Doroshkevich\ A.~G., Tucker\  D.~L., Allam\ S., \& Way\ M.~J.,  2004,   
\aap,\ 418, 7

       
\bibitem[Einasto et al.(1997a)]{einetal97a}
Einasto\ J., Einasto\ M., Frisch\ P.\ et al.,  
1997a,  \mnras,\ 289, 801 

 
\bibitem[Einasto et al.(1997b)]{einetal97b}
Einasto\ J., Einasto\ M., Frisch\ P.\  et al.,
1997b,  \mnras,\ 289, 813 

     
\bibitem[Einasto et al.(2001)]{einastetal01}
Einasto\ M., Einasto\ J.,  Tago\ E.,
M\"uller\ V., \& Andernach\ H.,\  2001,  \aj,\  122,  2222 

         
\bibitem[Eisenstein \& Hu(1998)]{eih98}
Eisenstein\ D.~J., \& Hu\ W., 1998,  \apj,\ 496, 605 

          
\bibitem[Eisenstein et al.(1998)]{eiht98}
Eisenstein\ D.~J., Hu\ W., \& Tegmark\ M., 1998, 
\apj,\ 504, L57 

   
\bibitem[Eisenstein et al.(2005)]{eisen05}
Eisenstein\ D.~J., Zehavi\ I., Hogg\ D.~W.\ et\ al., 2005, 
\apj,\ 633, 560 


\bibitem[Gazta\~{n}aga et al.(2009a)]{gch09}
Gazta\~{n}aga\ E., Cabr\'{e}\ A., \&  Hui\ L., 2009a,
\mnras, 399, 1663


\bibitem[Gazta\~{n}aga et al.(2009b)]{gms09}
Gazta\~{n}aga\ E., Miquel\ R., \& S\'{a}nchez\ E., 2009b,
Rhys.\ Rev.\ Lett., 103, 091302 


\bibitem[Hamilton(1993)]{ham93}
Hamilton\ A.~J.~S., 1993, \apj, 417, 19 


\bibitem[Harrison(1993)]{h93}
Harrison\ E., 1993, \apj, 403, 28 

  
\bibitem[Hartnett \& Hirano(2008)]{harhir08}
Hartnett\ J.~G., \& Hirano\ K., 2008, 
Ap\&SS,\ 318, 13 


\bibitem[Hartnett(2009)]{hart09}
Hartnett\ J.~G., 2009, 
Ap\&SS,\ 324, 13 

     
\bibitem[Hirano et al.(2008)]{hkk08}
Hirano\  K., Kawabata\ K., \& Komiya\ Z., 2008,
Ap\&SS, 315, 53 


\bibitem[Hogg(1999)]{h99} 
Hogg\ D.~W., 1999, preprint (astro-ph/9905116)


\bibitem[Junkkarinen et al.(1991)]{jhb91}
Junkkarinen\ V., Hewitt\ A., \& Burbidge\ G., 1991,
\apjs,\ 77, 203  
 
       
\bibitem[Kacprzak et al.(2007)]{kacp07}
Kacprzak\ G.~G., Churchill\ C.~W., Steidel\ C.~C., 
Murphy\ M.~T., \& Evans\ J.~L., 2007,
ApJ, 662, 909 

      
\bibitem[Kacprzak et al.(2008)]{kacp08}
Kacprzak\ G.~G., Churchill\ C.~W., Steidel\ C.~C., 
\&  Murphy\ M.~T., 2008,
AJ, 135, 922 
 
      
\bibitem[Kaiser \& Peacock(1991)]{kp91}
Kaiser\ N., \& Peacock\ J.~A., 1991, \apj,\ 379, 482 

       
\bibitem[Kayser et al.(1997)]{khs97}
Kayser\ R., Helbig\ P., \& Schramm\ T.,  1997,
\aap, 318, 680 

 
\bibitem[Kaminker et al.(2000)]{krv00}
Kaminker\ A.~D., Ryabinkov\ A.~I.,  \& Varshalovich\ D.~A., 2000,
\aap,\ 358, 1   
  

\bibitem[Kazin et al.(2010a)]{kaz10a}
Kazin\ E.~A., Blanton\ M.~R., Scoccimarro\ R.\  et al., 
2010a, \apj, 710, 1444 


\bibitem[Kazin et al.(2010b)]{kaz10b}
Kazin\ E.~A., Blanton\ M.~R., Scoccimarro\ R., 
McBridge\ C.~K., Berlind\ A.~A., 2010b,
preprint (arXiv:1004.2244)


\bibitem[Kerscher et al.(2000)]{kss00}
Kerscher\ M., Szapudi\ I., \& Szalay\ A.~S., 2000,
\apj, 535, L13 

        
\bibitem[Kurki-Suonio et al.(1990)]{ksmf90}
Kurki-Suonio\ H.,  Mathews\ G.~J., \& Fuller\ G.~M., 1990,
\apj, 356, L5 
 

\bibitem[Landy \& Szalay(1993)]{ls93}
Landy\ S.~D., \& Szalay\ A.~S., 1993, \apj,\ 412, 64 

 
\bibitem[Lundgren et al.(2009)]{lundetal09}
Lundgren\ B.~F., Brunner\ R.~J., York\ D.~G.\  et al.,\
2009, \apj, 698, 819

     
\bibitem[Mart\'{i}nez et al.(2009)]{martetal09}
Mart\'{i}nez\  V.~J.,  Arnalte-Mur\ P., Saar\ E.,\  et al.,\  
2009,\ \apj,\  696,\  L93;
Erratum: 2009,\ \apj,\ 703,\  L184

\bibitem[Mattig(1958)]{m58}
Mattig\ W., 1958,  Astron.\ Nachr., 284,  109
%

 
\bibitem[M\'{e}nard et al.(2008)]{menard08}
M\'{e}nard\ B., Nestor\  D., Turnshek\ D.,\  et al.,\ 
2008, \mnras, 385, 1053


\bibitem[Mo et al.(1992)]{mjb92}
Mo\ H.~J., Jing\ Y.~P., \& B\"orner\ G., 1992, \apj,\ 392, 452 

      
\bibitem[Morikawa(1991)]{m91}
Morikawa\ M., 1991,  \apj,\ 369,\  20


\bibitem[Peacock \& Nicholson(1991)]{pn91}
Peacock\ J.~A., \& Nicholson\ D., 1991, \mnras,\ 253, 307 


\bibitem[Peebles(1993)]{p93}
Peebles\ P.~J.~E., 1993, Principles of Physical Cosmology,
Princeton Univ. Press, Princeton


\bibitem[Percival et al.(2007a)]{percetal07a}
Percival\ W.~J., Cole\ S., Eisenstein\ D.~J.,\  et al.\  
2007a, \mnras,\ 381, 1053 

      
\bibitem[Percival et al.(2007b)]{percetal07b}
Percival\ W.~J., Nichol\ R.~C.,\ Eisenstein\ D.~J.\ et al.,
2007b, \apj,\ 657, 51 


\bibitem[Prochter et al.(2006)]{prprb06}
Prochter\ G.~E., Prochaska\ J.~X., \& Burles\ S.~M., 
2006, \apj, 639, 766


\bibitem[Quashnock et al.(1996)]{qby96}
Quashnock\ J.~M.,  Vanden Berk\ D.~E., \& York\ D.~G.,  1996,
\apj,\  472, L69  


\bibitem[Rivolo(1986)]{r86}
Rivolo\ A.~R., 1986,\ \apj,\  301, 70

  
\bibitem[Ryabinkov et al.(2003)]{rkv03}
Ryabinkov\ A.~I., Kaminker\ A.~D., \&  Varshalovich\ D.~A., 2003,
\aap,\ 412, 707; \\  
www.ioffe.ru/astro/QC     


\bibitem[Ryabinkov et al.(2007)]{rkv07}
Ryabinkov\ A.~I., Kaminker\ A.~D., \& Varshalovich\ D.~A., 2007,
\mnras,\ 376, 1838\ (Paper~I)  



\bibitem[Tago et al.(2002)]{tagoetal02}
Tago\ E., Saar\ E., Einasto\ J.\ et al., 
2002, \aj,\ 123, 37 

        
\bibitem[van de Weygaert(1991)]{vdw91}
van de Weygaert\ R.,  1991, \mnras, 249, 159 

  
\bibitem[Yoshida et al.(2001)]{yosh01}
Yoshida\ N., Colberg\ J., White\ S.~D.~M.\ et al.,
2001, \mnras,\ 325, 803
 

\end{thebibliography}
\end{document}